\documentclass{article}
\usepackage{a4wide}
\usepackage{graphicx}
\usepackage{amsmath}
\usepackage{amssymb}
\newcommand{\IR}{\mathbb{R}}
\newcommand{\TM}{\textsf{TM}}
\newcommand{\TMs}{\textsf{TM}s}
\newcommand{\person}[1]{\textsc{#1}}
\newcommand{\IN}{\mathbb{N}}
\newcommand{\Life}{\textsf{Life}}
\newcommand{\alive}{{\tt alive}}
\newcommand{\dead}{{\tt dead}}
\newcommand{\Java}{\textsf{Java}}
\newcommand{\COMMENTED}[1]{}
\newtheorem{theorem}{Theorem}[section]
\newtheorem{remark}[theorem]{Remark}
\newtheorem{observation}[theorem]{Observation}
\newtheorem{definition}[theorem]{Definition}
\newtheorem{lemma}[theorem]{Lemma}
\newtheorem{corollary}[theorem]{Corollary}
\DeclareRobustCommand{\qed}{%
\ifmmode 
\else \leavevmode\unskip\penalty9999 \hbox{}\nobreak\hfill
\fi
\quad\hbox{\leavevmode%
\hbox to.77778em{%
\hfil\vrule
\vbox to.675em{\hrule width.6em\vfil\hrule}%
\vrule\hfil}}}

\makeatletter
\newenvironment{proof}{\par
\normalfont
\topsep6\p@\@plus6\p@ \trivlist
\item[\hskip\labelsep\itshape
\bf Proof\@addpunct{:}]\ignorespaces
}{%
\qed\endtrivlist
}
\begin{document}
\title{
Does Quantum Mechanics allow for Infinite Parallelism?}
\author{Martin Ziegler, University of Paderborn, 33095 GERMANY}
\date{
}
\maketitle
\begin{abstract}
Recent works have independently suggested that Quantum
Mechanics might permit for procedures that
transcend the power of Turing Machines
as well as of `standard' Quantum Computers.
These approaches rely on and indicate that
Quantum Mechanics seems to support some infinite
variant of classical parallel computing.

We compare this new one with other attempts towards
hypercomputation by separating 1)~its principal
computing capabilities from 2)~realizability issues. The
first are shown to coincide with recursive enumerability;
the second are considered in analogy to
`existence' in mathematical logic.
\end{abstract}
{\textbf{Keywords}:
  Hypercomputation, Quantum Mechanics, Recursion Theory,
  Infinite Parallelism. }
\section{Computability} \label{secComputability}
In 1936, \person{Alan M. Turing} gave an example of a
well-founded (and thus mathematically solvable) problem
which he showed to admit no computable solution.
More precisely he first introduced what is now called the
Turing Machine (\TM) as an idealization (`\emph{model}') 
of a digital
computer and revealed it capable of solving a vast variety
of practical problems like, for instance, answering
whether a given integer is prime.
\begin{definition} \label{defProblem}
A computational \emph{problem} is a subset $L\subseteq\IN$. \\
It is \emph{decided} by \TM{} $M$ if,
upon input of any $x\in\IN$,
\vspace*{-1ex}\begin{itemize}\itemsep0pt
\item $M$ eventually outputs ``0'' (\emph{rejects}) and halts
  ~in case $x\not\in L$.
\item $M$ eventually outputs ``1'' (\emph{accepts}) and halts
  ~in case $x\in L$
\end{itemize}
\end{definition}
Having thus indicated the fundamental power of this machine,
\person{Turing} then proceeded to exhibit its limitation by
formally proving that the \textsf{Halting Problem} $H$
--- the question whether a given \TM{} $M$ eventually
halts or rather continues executing indefinitely ---
cannot be decided by any \TM $M_0$.
Notice that, according to Definition~\ref{defProblem},
this hypothetical $M_0$ is required to always give the
correct answer and to terminate.
More precisely, the difficulty inherent to the Halting Problem
consists in telling within finite time whether $M$
does \emph{not} halt; for, simply simulating
$M$ step by step, $M_0$ can easily identify the case
when $M$ \emph{does} terminate.

\person{Turing}'s result initiated the flourishing
field of Computability or Recursion Theory \cite{Odifreddi}.
Its goal is to distinguish computable from uncomputable
problems and to classify the latter according to their degree
of uncomputability \cite{Soare}. For example, the following celebrated
result of \person{Matiyasevich} has settled \person{Hilbert}'s
\textsf{Tenth Problem} to the negative by proving it
equivalent to $H$:
\begin{theorem}[\cite{Matiyasevich}] \label{thHilbert}
On the one hand, a given description of a Diophantine Equation
$E$ (like \person{Fermat}'s famous ~``$a^n+b^n=c^n$''~
for $a,b,c,n\in\IN$) can computably
be transformed into a \TM{} $M$ such that it holds: $M$
terminates iff $E$ admits an integral solution.

On the other hand, a given \TM{} $M$ can computably be converted
into the description of a Diophantine Equation $E$ in
such a way that, again, $M$ halts iff $E$ admits an integral solution.
In particular since, according to \cite{Turing}, the first
cannot be decided algorithmically, neither can the latter.
\end{theorem}
Apart from the \TM{}, many further sensible notions of
computability have been proposed: e.g., $\mu$-Recursion
(which gave the field its name),
\person{Herbrand}-\person{G\"{o}del}-Computability
(which led to the programming language \texttt{Prolog}),
or $\lambda$-Calculus (which stipulated \texttt{Lisp}).
But they were all shown equivalent to the \TM{} by
\person{Church}, \person{Kleene}, \person{Post}, \person{Turing} 
and others; cf. e.g. \cite[\texttt{Chapter~I}]{Odifreddi}
or \cite[\texttt{Section~26.3+4}]{Atallah}.
The popular language \Java{} for example is equivalent, too:
internet provides many applets for simulating a
\TM{}; and conversely devising
a \Java{} interpreter on a \TM.
is merely tedious but not difficult.
Even a \textsf{Pentium}\textregistered~ processor
is basically just a \TM --- although a very fast one;
recall that we are dealing with problems which cannot
be solved computationally at all, neither quickly nor slowly.
For the very same reason, (at least `standard') Quantum Computers
are still no more powerful than an ordinary \TM{}
\cite[p.3 footnote~1]{QC}.

\subsection{Church-Turing Hypothesis}
It should be emphasized that both Halting and
{Hilbert}'s Tenth Problem are desirable
to be solved for very practical reasons.
The first for instance arises in automatized software
verification; indeed, correctness of
some \Java{} source code includes its termination
which, by the above considerations, cannot be checked
algorithmically.
Similarly, a hypothetical algorithm for deciding feasibility of
Diophantine Equations could be applied to computer-proving
not only \person{Fermat}'s Last Theorem but also to settle many
other still open questions for example in number theory.

Observing that \cite{Turing,Matiyasevich} ruled out the
possibility of a \emph{Turing Machine} to decide either of these problems,
people have since long tried to devise other computing devices
exceeding its principal power. However the perpetual
failure to do so\COMMENTED{\footnote{compare also
\cite[\textsc{Chapter~10}]{Adams}:
``\emph{Many respectable physicists [...]
couldn't stand [...] the perpetual failure they encountered
in trying to construct a machine which could generate the
\emph{infinite} improbability field needed [...],
and in the end they grumpily announced that such a machine
was virtually impossible.
\\
Then, one day, a student who had
been left to sweep up the lab after a particularly unsuccessful
party found himself reasoning this way:
If, he thought to himself, such a machine is \emph{virtual}
impossible, then it must logically be a \emph{finite} improbability.
So all I have to do in order to make one is to work out
exactly how improbable it is, feed that figure into the finite
improbability generator, give it a fresh cup of really hot tea
\ldots and turn it on!
He did this, and was rather startled to discover that he
had managed to create the long-sought-after golden
Infinite Improbability generator out of thin air.\\
It startled him even more when just after he was awarded the
Galactic Institute's Prize for Extreme Cleverness he got lynched
by a rampaging mob of respectable physicists who had finally
realized that the one thing they really couldn't stand was a smart-ass.%
}''}}
plus the aforementioned results of a \TM{}
being able to simulate many other notions of computability
have eventually led to what has become known as the
\textsf{Church-Turing Hypothesis}:
\begin{quote} \it Anything that can be computed in practice,
is also computable by a \TM{}. \end{quote}

We emphasize that, due to its informal nature,
this hypothesis cannot be proven formally. Informal
arguments in its favor usually point out that
computation is a physical process which, by mathematically
describing the physical laws it is governed by, can be
simulated by a \TM{} up to arbitrary finite numerical precision;
and infinite accuracy were required only for `chaotic' processes
which are too sensitive to perturbations than
being harnessable for practical computation anyway.

However it has later been pointed out that certain theories
of quantum gravitation
might actually \emph{not} admit a simulation by
a \TM{} \cite[p.546]{Geroch}; furthermore even Classical Mechanics
seems to conceivably provide for processes whose simulation
requires infinite precision during \emph{intermediate} times only,
whereas the resulting behavior is asymptotically stable and thus
suited well to realize a non-Turing form of physical computation
\cite{Yao}. Moreover the laws of nature we know so far cannot
be deduced to fundamentally restrict computation \cite{Bennett}.

In fact neither \person{Church} nor \person{Turing}
themselves have put forward a claim as universal as the
way `their' hypothesis is often (mis-)interpreted \cite{CopelandCT}.
Instead, literature contains and discusses a rich variety of
related hypotheses \cite[\textsc{Section~2.2}]{Ord}.

\subsection{Hypercomputation}
Anyway, the question remains open whether there might
exist a computing device more powerful than the \TM{} or not.

\begin{remark} \label{remPost}
Already \person{Post} wondered whether \emph{any}
super-\TM{} has the ability to solve the Halting Problem.
This was settled in the negative by
\person{Friedberg} and \person{Muchnik} independently
showing that there exists an entire hierarchy of undecidable
problems strictly `easier' than $H$.
More precisely they constructed problems $P\subseteq\IN$
that a \TM{} can\emph{not} decide, yet access to
whose solution (in terms of an oracle, that is,
by permitting queries
to some hypothetical external device answering
questions ``$y\in P$'') provably still does
not enable this \emph{super}-\TM{} to solve the Halting Problem.
However, in contrast to $H$, such $P$ seem to be artificial.
See \cite[\textsc{Sections~5} to 7]{Soare} for a
more detailed account of \person{Post}'s Problem.
\end{remark}

To get an idea towards how a \emph{Hyper}computer
might look like, Theoreticians have started
considering super-\TMs{} and their
respective fundamental computing capabilities.
This established the flourishing
field of research called `{Hypercomputation}'
\cite{Proudfoot} which entire volumes of 
significant journals 
have become dedicated to \cite{UMC,TCS317}.
Devising such a formal model (i.e., an idealized abstraction)
of a Hypercomputer proceeds in many cases
less by adding extensions to but rather by
removing restrictions from a \TM{}.
\begin{observation} \label{obsFinite}
The \TM{} is characterized by
\vspace*{-1ex}\begin{enumerate}\itemsep0pt
\item[a)] a finite control (the `program', so to speak);
\item[b)] an initially blank, countable supply of memory cells
\item[c)] storing a finite amount of information each
  (e.g., a bit or an integer);
\item[d)] finite running time;
\item[e)] possibly finite parallelism (as, e.g., for
  a \emph{nondeterministic} \TM).
\end{enumerate}\vspace*{-1ex}
Irregardless of the details of its precise definition,
these finiteness conditions directly imply
that there is an at most countable number
of different \TMs; whereas computational problems
according to Definition~\ref{defProblem} exist of
cardinality 
of the continuum. Thus, most of them are undecidable.
\end{observation}

Conditions~a)-e) underlie the mourned
limitations of the classical \TM, and dropping one or more
of them leads to several well-known models of hypercomputation;
see, e.g. \cite[\textsc{Section~3}]{Ord}
or \cite[\textsc{Section~2}]{CopelandSurvey}.
\emph{Oracle} Machines for instance,
subject of \person{Turing}'s Dissertation in Princeton \cite{Turing3}
and now core of Recursion Theory \cite{Soare,Odifreddi},
correspond to \TMs{} with initial memory inscription,
that is, they remove Condition~b);
\person{Blum, Shub, and Smale}'s $\IR$-Machine
\cite{BSS} abolishes Condition~c) by allowing
each cell to store a real number;
while \emph{Infinite Time} Machines due to
\cite{Hamkins} lift Condition~d).

The proposal, consideration, and investigation of such enhanced
abstract models of computation and their respective computational
powers by Logicians and Theoretical Computer Scientists has
proven particular seminal regarding related
contributions from Theoretical Physics on their realizability.
For example, \cite{Tucker} has indicated that a physical system
breaking Condition~a) might actually
exist\footnotemark{}\addtocounter{footnote}{-1};
while \cite{Hogarth,Shagrir} pointed out that
in \textsf{General Relativity} there might
exist\footnote{refer to Remark~\ref{remExist} below}
space-time geometries allowing to watch within
finite time a computer execute an infinite number
of steps and thus to lift Condition~d).

\subsection{Quantum Mechanical Hypercomputation} \label{secNew}
Recently, several new approaches have been
suggested for solving either the Halting Problem
\cite{Svozil1,QuantumCoins,Calude}
or {Hilbert}'s Tenth Problem
\cite{KieuCP,KieuIJTP}. They exploit Quantum
Mechanics and thus form a nice counterpart to
previous approaches based on General Relativity
\cite{Hogarth,Shagrir} as the other pillar of
non-classical physics. Recalling that `standard'
Quantum Computing does {not} exceed 
Turing's Barrier, these approaches must be 
non-standard in some sense which closer 
inspection reveals to be infinite parallelism:
\vspace*{-1ex}\begin{itemize}\itemsep0pt
\item
``\emph{Our quantum algorithm is based on [\ldots]
  our ability to implement physically certain Hamiltonians
 having infinite numbers of energy levels}''
\qquad\qquad\qquad\quad\cite[top of \textsc{Section~6.3}]{KieuCP};
\item
``\emph{The key ingredients are the availability of
a countably infinite number of Fock states, the ability
to construct/simulate a suitable Hamiltonian}''
 \hfill\cite[end of \textsc{Section~4}]{KieuIJTP};
\item
``\emph{The new ingredients built in our `device' include the
use of an infinite superposition (in an infinite-dimensional
Hilbert space) which creates an `infinite type of
quantum parallelism'}\,''  \hspace*{\fill}
\cite[p.123 \textsc{Section~5}]{QuantumCoins}.
\end{itemize}\vspace*{-1ex}

\noindent
It thus seems that Quantum Mechanics allows to drop
Condition~e) from Observation~\ref{obsFinite} and so
to provide a new promising approach to the
existence\addtocounter{footnote}{-1}\footnotemark{}
of hypercomputers --- an approach \emph{not} covered by
\person{Ord}'s classification \cite[\textsc{Section~3}]{Ord}.
The present work describes in Section~\ref{secParallel}
the theoretical consequences from lifting Condition~e),
that is the principal computing power of infinite parallelism.

We conclude this section with an already announced remark
on the notion of existence.
\begin{remark} \label{remExist}
The question whether a physical device with certain properties
\emph{exists} bears logical similarity to the question
whether a mathematical object with certain properties
exists.
\noindent
In the latter case for a proof, only very few
(namely constructive or
intuitionistic) mathematicians will
\vspace*{-1ex}\begin{enumerate}\itemsep0pt
\item[A)] insist that one
actually \emph{constructs} this object
\end{enumerate}\vspace*{-1ex}
whereas most
are satisfied for instance with
\vspace*{-1ex}\begin{enumerate}\itemsep0pt
\item[B)] an indirect argument
showing that its \emph{non}-existence leads to a
contradiction.
\end{enumerate}\vspace*{-1ex}
In fact a majority of contemporary mathematicians 
will even take it for granted if
\vspace*{-1ex}\begin{enumerate}\itemsep0pt
\item[C)] the object's \emph{existence} does \emph{not}
lead to a contradiction.
\end{enumerate}\vspace*{-1ex}
For example the claim
``\emph{To every vector space there exists a basis}''
is of kind~C) as well as many principles
in \textsf{Functional Analysis}: Each of them
is equivalent to the \textsf{Axiom of Choice}
\cite{BasisAC}
and thus does not lead to a contradiction to
conservative set theory (C) but cannot be
deduced from it (B) as has first been proven
by \person{K.~G\"{o}del} in \cite{Goedel}
and later strengthened by \person{P.J.~Cohen}.

Similarly, the existence of a {physical} object can
be proven in a strong way A) by actually constructing ist.
But in most cases, showing it C) consistent with physical
laws is accepted as well. Observe that this is how
both Positrons as well as Black Holes came to existence:
as solutions of (and thus consistent with)
\person{Dirac}'s Equation and \person{Einstein}'s
General Relativity, respectively; only later have
new experimental observations upgraded
their existence to type B).
\end{remark}
\section{Infinite Parallel Computing} \label{secParallel}
The prospering field of Parallel Computing knows and has agreed
upon a small collection of models as theoretical abstractions for
devising and analyzing new algorithms for various actual parallel
machines \cite[\textsc{Sections~45.2} and \textsc{47.2}]{Atallah}.
Of course with respect to their principal
power, that is computability rather than complexity,
they are all equivalent to the \TM.

However when talking about \emph{infinite} parallelism,
seemingly no such agreement has been established,
cf. e.g. \cite[p.284]{Eberbach}; and
in fact no equivalence, either, as will turn out.
For instance, of what kind are the countably infinitely
many individual computers that are to operate concurrently ---
\TMs{} or finite automata? In the first case,
do they all execute the same program? When 
is the result to be read off? The answers to these
questions fundamentally affect the capabilities
of the resulting system.
\subsection{Infinite Cellular Automata}
Consider parallelism in an infinite cellular
automaton in the plane. More specifically, we refer
to \person{Conway}'s famous \textsf{Game of Life}
\cite[\textsc{Ch.~25}]{cellular}
where in each step, any cell's successor state concurrently
is determined by its present state as well as those of
its eight adjacent ones' as follows
(cf. Figure~1):
\vspace*{-1ex}\begin{itemize}\itemsep0pt
\item A \dead{} cell with exactly three neighbors \alive{}
  becomes \alive, too; \\ otherwise it remains \dead.
\item A living cell with two or three neighbors \alive{}
  stays \alive; \\ otherwise (0,1,4\ldots 8 living neighbors,
  that is) it dies.
\end{itemize}\vspace*{-1ex}
\begin{center}
\includegraphics[width=0.27\textwidth]{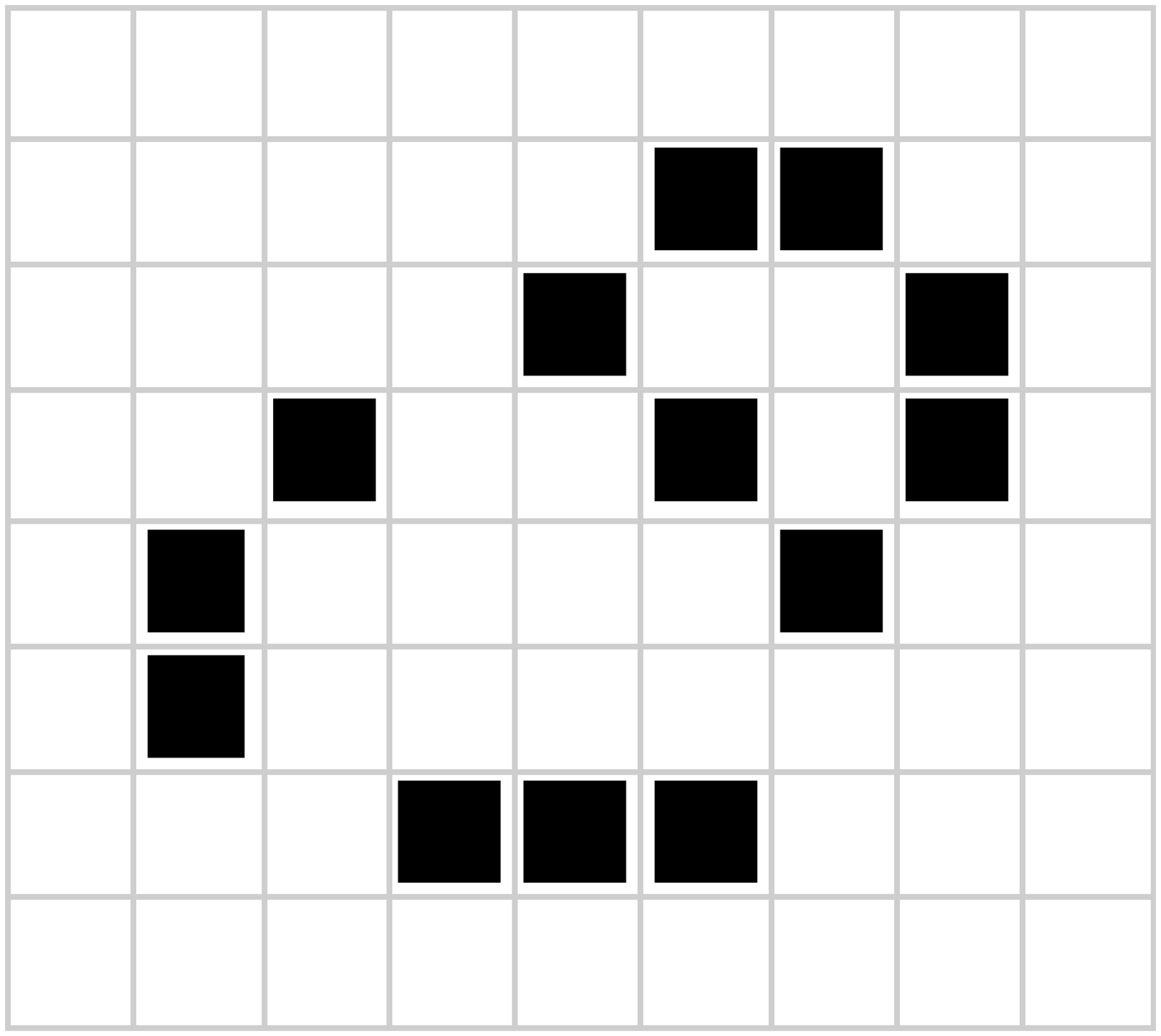}
\hfill
\includegraphics[width=0.27\textwidth]{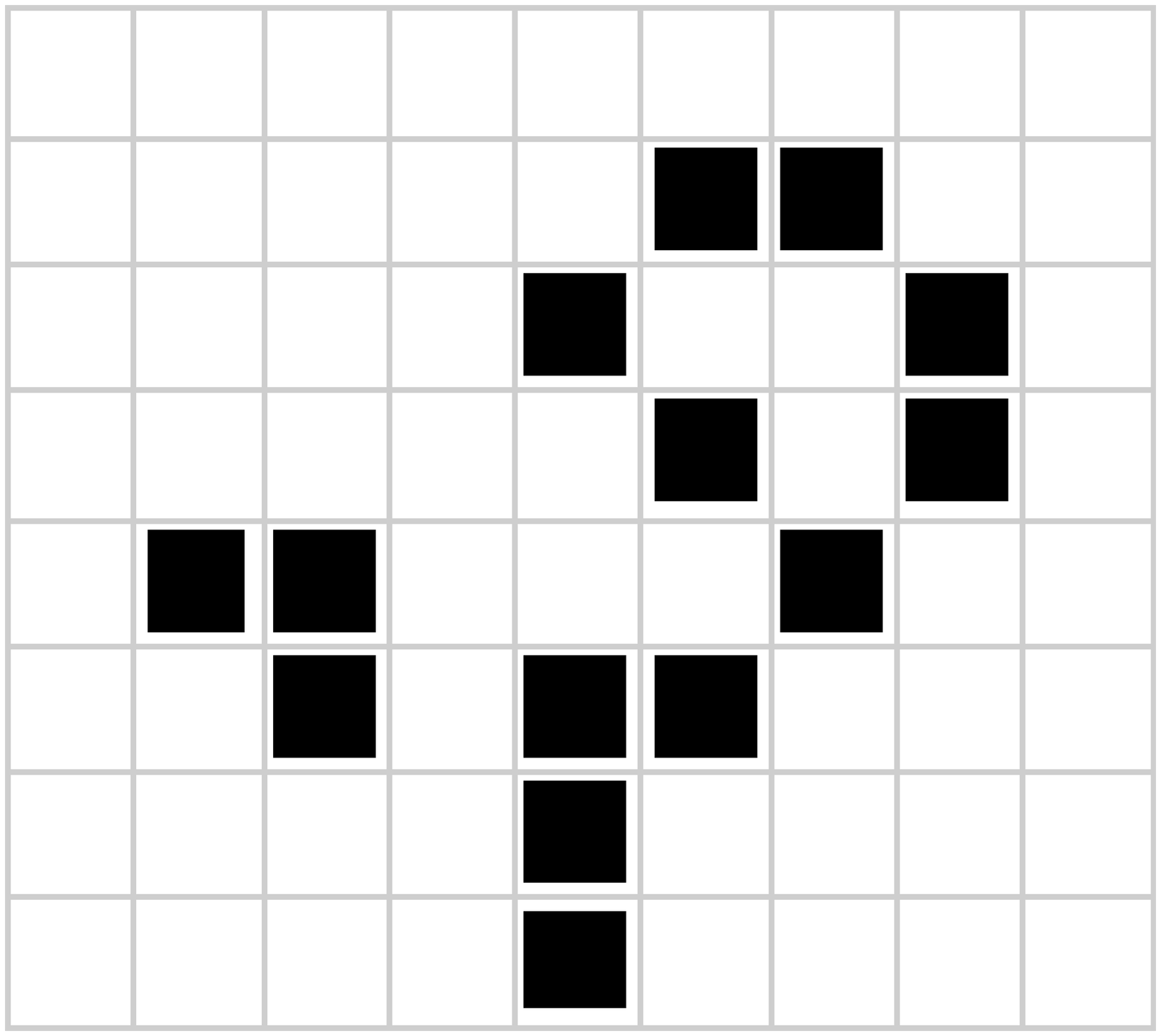}
\hfill
\includegraphics[width=0.27\textwidth]{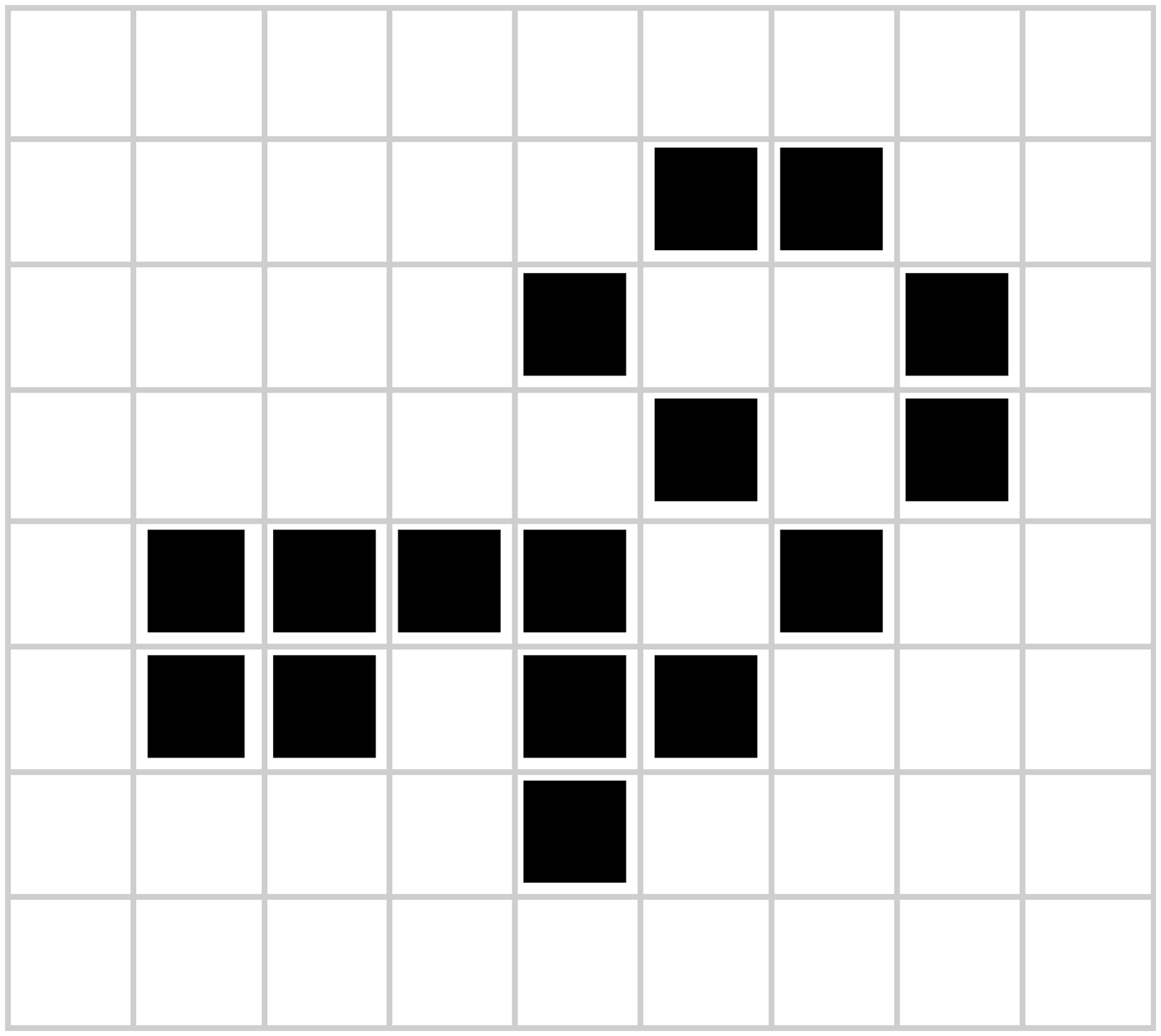}
\\[0.7ex]{Figure~1: \small\sc Sample Initial and Two Successor
Configurations of} \Life.
\end{center}
\begin{definition} \label{defLife}
Starting from a given initial configuration, \Life{}
\emph{terminates} if the sequence of successor configurations
eventually stabilizes.
This resulting configuration is called
\emph{rejecting} if it is empty (every cell \dead),
otherwise \emph{accepting}.
\\
An initial configuration is \emph{finite} if,
out of the infinite number of cells, only
finitely many are \alive.
\end{definition}
\begin{theorem} \label{thLife}
\Life{} with finite initial configuration is
a system with infinite parallelism yet
equivalent to the \TM. More precisely:
\vspace*{-1ex}\begin{enumerate}\itemsep0pt
\item[i)]
  Given a finite initial configuration,
  its evolution through \Life{} can be simulated by
  a \TM.
\item[ii)]
  There is a finite initial configuration capable
  of simulating the Universal (and thus any) \TM.
\end{enumerate}\vspace*{-1ex}
\emph{Simulation} here means:
The \TM{} terminates/accepts/rejects ~iff~ \Life{} 
terminates/accepts/rejects.
\end{theorem}
\begin{proof}
The internet provides plenty implementations for \Life{}
written, e.g., in \Java. Feeding one to the aforementioned
\Java{} interpreter for \TMs{} yields i). 
Claim~ii) is in fact a famous result
based on an ingenious and complicated construction;
refer to
\cite[\textsc{Ch.~25}]{cellular}
for details.
\end{proof}
In particular, \Life{} matches but does \emph{not}
exceed the computing capabilities of a \TM{};
cf. \cite[\textsc{Section~26.4}(1)]{Atallah}.

Here, finiteness of the initial configuration
enters crucially of course. One can indeed show that
\emph{in}finite initial configurations in ii) correspond
to non-blank memory contents and thus to dropping
in Observation~\ref{obsFinite} both Conditions~e) \emph{and} b).
\subsection{Infinite Turing Concurrency}
In order to focus on the power obtained from infinite parallelism
\emph{only} (that is, by removing just Condition~e),
now consider the finite automata replaced by \TMs.
Indeed the three citations in Section~\ref{secNew}
indicate that, whereas Quantum Computing --- and
in particular simulating a single classical \TM{} ---
requires only a finite (or at most countably infinite)
number of dimensions
from quantum mechanical Hilbert Space, its
\emph{in}finitely many dimensions provide
room for an infinite number of \TMs:
cf. \textsf{Hilbert's Hotel}.

Strictness of \person{Chomsky}'s Hierarchy
implies that a single \TM{} is provably more
powerful than a single automaton
\cite[\textsc{Section~25.3}]{Atallah}.
One may therefore expect that the capabilities of
an infinite number of \TMs{} exceed those of an
infinite number of automata (and thus actually
lead to hypercomputation); by how much, however,
turns out to depend.

In analogy to Definition~\ref{defProblem}, consider
first the following notion of solving a problem
by means of infinite parallelism.
\begin{definition} \label{defParallel}
Fix a problem $L\subseteq\IN$
and a countably infinite family $(M_k)_{_{k\in\IN}}$ of \TMs.
This family \emph{solves} $L$ if,
for each $x\in\IN$,
\vspace*{-1ex}\begin{enumerate}\itemsep0pt
\item[i)] each $M_k$, upon input $x$, eventually terminates \qquad and
\item[ii)] at least one $M_k$ outputs ``1'' (accepts) ~iff~ $x\in L$.
\end{enumerate}
\end{definition}
However observe that, whereas each individual $M_k$ halts,
the time required to do so may depend on $k$ so
that it takes infinitely long for the entire
family $(M_k)_{_k}$ to terminate.
(We point out that this behavior resembles the
\emph{fair infinite nondeterminism} of \cite{Boas}.)
In order to know the result within finite time,
the following additional
requirement is therefore important:

\smallskip
\smallskip
\noindent {\bf Definition~\ref{defParallel} (continued)~}
{\it\vspace*{-1ex}\begin{enumerate}\itemsep0pt
\item[iii)] if upon input of any $x\in\IN$,
  all $M_k$ terminate within finite time bounded independently of $k$.
\end{enumerate}}
While seeming sensible at first glance, a second
thought reveals that, even with this restriction,
the resulting notion of `infinitely parallel computability'
is still unreasonable: simply because
\emph{any} problem $L\subseteq\IN$ becomes trivially
solvable by an appropriate family $(M_k)_{_k}$.
To this end let the program executed by $M_k$
store the constant ``1'' if $k\in L$
and the constant ``0'' otherwise.
Let its main part then operate as follows:
Upon input of $x\in\IN$
test whether $x=k$;
if so, output the stored constant,
otherwise output ``0''; then terminate.
\\
The point is of course that the according family $(M_k)_{_k}$
solving $L$ is only shown to \emph{exist}. More precisely
there is in general no means of \emph{computing}, given
$k\in\IN$, a description of $M_k$ and its constants.
This insight suggests to finally add the following
further requirement, in Theoretical Computer Science
known as a \emph{uniformity} condition.

\smallskip
\smallskip
\noindent{\bf Definition~\ref{defParallel} (concluded)~}
{\it\vspace*{-1ex}\begin{enumerate}\itemsep0pt
\item[iv)] if a \TM{} $M_0$ is capable of generating, upon input of $k$,
  (the encoding of) ~$M_k$.
\end{enumerate}}
\noindent
Here, \emph{encoding} refers to a sort of `{blueprint}' of $M_k$
or, more formally, its \person{G\"{o}del} Number
\cite[\textsc{Section~9.1.2}]{Hopcroft}.
\subsection{Computational Power of Infinite Turing Concurrency}
This section reveals that the Definition~\ref{defParallel}(i-iv)
indeed yields an interesting non-trivial way of hypercomputation.
More precisely we show that, in this sense, infinite Turing-Parallelism
can
\vspace*{-1ex}\begin{itemize}\itemsep0pt
\item solve the Halting Problem $H$
\item as well as {Hilbert}'s Tenth Problem
\item but \emph{not} \textsf{Totality}.
\end{itemize}\vspace*{-1ex}
While $H$ refers to the question whether given \TM{} $M$,
started on a single given input $x$, eventually terminates,
\textsf{Totality} asks whether $M$ halts on \emph{all}
possible inputs. So in contrast to the first, this
even more important property of correct software still remains
intractable to automated checking even on this kind
of hypercomputer.

\begin{theorem} \label{thH}
The Halting Problem is solvable by an infinity of
\TMs{} working in parallel in the sense of Definition~\ref{defParallel}(i-iv).
\end{theorem}
\begin{proof}
For each $k\in\IN$, let $M'_k$ proceed as follows:
Given $M$ and $x$, simulate the first $k$ steps of $M$ operating on $x$;
if $M$ halts within these steps, then output ``1'' and terminate;
otherwise output ``0'' and terminate.
\\
Observe that the family $(M'_k)_{_k}$ satisfies
i) and ii) from Definition~\ref{defParallel}. Moreover,
one easily confirms (iv) that an appropriate \TM{} $M'_0$
can indeed generate from $k$ an encoding of this $M'_k$.
Based on the Universal \TM{} $U$,
$M'_k$ can be achieved to have running time
$t(n)\leq c\cdot(n\cdot k)^2$ for some constant $c$;
combine for example 
\cite[the \textsc{Lemma} in \textsc{Section~4}]{Fuerer}
with \cite[\textsc{Theorem~8.10}]{Hopcroft}.
Here, $n=|x|+|M|$ denotes the joint length of the 
binary encoding of $x$ and $M$.
\\
Now let $M_k$ be the \TM{} obtained from applying
the below \emph{Linear Speed-Up Lemma}~\ref{lemSpeedup}
to $M'_k$ with $C:=k^3$.
It follows that $M_k$ has running time independent of $k$,
that is, it does comply with iii) while still satisfying
i), ii), and iv).
\end{proof}

In order to achieve Property~iii) in the above proof,
the crucial ingredient is the below
classical construction. It basically says that any \TM{}
can be speed up by a \emph{constant} factor.
\begin{lemma}[Linear Speed-Up] \label{lemSpeedup}
For each $C\in\IN$ and any \TM{} $M'$ of time complexity $t(n)$,
there exists another \TM{} $M$ simulating $M'$
within running time $n+t(n)/C$.
\\
$M$ can be obtained computationally from
$M'$; i.e., there is a fixed further \TM{} which,
given an encoding of any $M'$ and $C$,
outputs an encoding of $M$ as above.
\end{lemma}
\begin{proof}
See for instance \cite[\textsc{Theorem~24.5}(b)]{Atallah}.
\end{proof}

The solvability of Hilbert's Tenth Problem
now follows from Theorem~\ref{thHilbert} and
Theorem~\ref{thH}.
More generally, the latter implies that
infinite Turing-Parallelism can solve
\emph{any} semi-decidable Problem $L\subseteq\IN$.
In fact, the converse holds as well:
\begin{theorem} \label{thRE}
A Problem $L\subseteq\IN$ is solvable in the sense of
Definition~\ref{defParallel}(i-iv) ~ iff~ 
semi-decidable.
\end{theorem}
Recall that `\emph{semi}-decidability'
(also called \emph{recursive enumerability})
weakens `decidability' from
Definition~\ref{defProblem} in that, here,
the \TM{} is allowed in case $x\not\in L$
to not halt but to loop endlessly \cite[\textsc{Section~825}]{Hopcroft}.
\begin{proof}
By the above remark, it remains to consider the case that
$L$ is solvable by some parallel family $(M_k)_{_k}$ according
to Definition~\ref{defParallel}. \\
Upon input of $x\in\IN$, sequentially simulate the $(M_k)_{_k}$ as follows:
\quad For each $k\in\IN$,
\vspace*{-1ex}\begin{itemize}\itemsep0pt
\item obtain from $M_0$ a description of $M_k$ ~by virtue of iv)
\item and simulate $M_k$ on input $x$.
  \quad (Observe its termination according to Property~i)
\item If output is ``1'', halt;
  ~otherwise proceed with next $k$.
\end{itemize}\vspace*{-1ex}
This algorithm indeed terminates iff
at least one $M_k$ outputs ``1'',
that is (ii), ~iff~ $x\in L$.
\end{proof}

\begin{corollary} \label{thTotality}
Even infinite Turing concurrency 
in the sense of Definition~\ref{defParallel}(i-iv)
cannot solve \textsf{Totality}.
\end{corollary}
\begin{proof}
\textsf{Totality} is well-known to \emph{not} be
semi-decidable. More specifically, we refer to
\cite[\textsc{Theorem~IV.3.2}]{Soare}
where this problem is shown to be
$\Pi_2$-complete, that is,
\cite[\textsc{Definition~IV.2.1} and \textsc{Corollary~IV.2.2}]{Soare}
reducible to $\overline{\emptyset^{(2)}}\not\in\Sigma_2$,
and therefore does not belong to the class
$\Sigma_1\subseteq\Sigma_2$
of recursively enumerable problems.
\end{proof}
\section{Conclusion}
Section~\ref{secComputability} has
pointed out that recent and independent
approaches due to
\person{Kieu}, \person{Calude}, and
\person{Pavlov}
to hypercomputation
via quantum mechanics
rely on some sort of infinite parallelism.
Regarding the respective complicated intertwined
quantum mechanical
constructions, procedures, and analyses,
we suggest to bring more clarity into this
subject by considering
algorithmic/ computational issues separately from
physical ones. This leads to
the following two questions to be treated individually:
\vspace*{-1ex}\begin{itemize}\itemsep0pt
\item[1)]
Does Quantum Mechanics allow for infinite parallelism;
and, if so, of what kind?
\item[2)]
What kinds of 
idealized infinite parallelism yield which principal computational
power; that is, does it and by how far
exceed the fundamental capabilities of a \TM{}?
\end{itemize}\vspace*{-1ex}
Section~\ref{secParallel} contains answers to the second question.
It reveals that in fact infinite
\emph{classical} (i.e., Turing-) parallelism
is sufficient for solving both the Halting Problem
as well as Hilbert's Tenth Problem.
This leaves open whether the infinite dimensions
of quantum mechanical Hilbert Space do indeed allow for this kind
of infinite classical parallelism. Specifically,
\vspace*{-1.3ex}\begin{itemize}\itemsep-2pt
\item[\labelitemii] preparation of a certain initial state,
\item[\labelitemii] its maintenance (in particular coherence) 
  through-out the computational evolution,
\hfill and
\item[\labelitemii] read-out of the final result
\end{itemize}\vspace*{-1.3ex}
are likely to raise here even more difficulties than already in the
finite-dimensional case of `standard' Quantum Computing
\cite[\textsc{Section~7.2}]{QC}.
For example only recently has it become possible
to read out a single spin \cite{Readout}.
However 
(im-)practicality of hypercomputation should 
not be confused with (un-)existence,
particularly in the light of Remark~\ref{remExist}.

\end{document}